\documentclass[aps,pra,twocolumn,superscriptaddress,showpacs]{revtex4}
\usepackage{tipa}
\usepackage{pifont}
\usepackage{txfonts}
\usepackage{amssymb}
\usepackage{graphicx}

\begin{document}

\title{Multiband effects on the conductivity for a multiband Hubbard model}
\author{Shi-Jian Gu}
\email{sjgu@phy.cuhk.edu.hk} \affiliation{Department of Physics and ITP, The
Chinese University of Hong Kong, Hong Kong, China}
\author{Junpeng Cao}
\affiliation{Beijing National Laboratory for Condensed Matter Physics, Institute of
Physics, Chinese Academy of Sciences, Beijing 100190, China}
\author{Shu Chen}
\affiliation{Beijing National Laboratory for Condensed Matter Physics, Institute of
Physics, Chinese Academy of Sciences, Beijing 100190, China}
\author{Hai-Qing Lin}
\affiliation{Department of Physics and ITP, The Chinese University of Hong Kong, Hong
Kong, China}

\begin{abstract}
The newly discovered iron-based superconductors have attracted lots of
interests, and the corresponding theoretical studies suggest that the system
should have six bands. In this paper, we study the multiband effects on the
conductivity based on the exact solutions of one-dimensional two-band Hubbard
model. We find that the orbital degree of freedom might enhance the critical
value $U_c$ of on-site interaction of the transition from a metal to an
insulator. This observation is helpful to understand why undoped High-$T_c$
superconductors are usually insulators, while recently discovered iron-based
superconductors are metal. Our results imply that the orbital degree of freedom
in the latter cases might play an essential role.
\end{abstract}

\date{\today}
\maketitle

%71.30.+h Metal¨Cinsulator transitions and other electronic transitions
%74.70.-b Superconducting materials (for cuprates, see 74.72.?h)
%71.10.Fd Lattice fermion models (Hubbard model, etc.)

%\section{Introduction}

Recently discovered iron-based superconductors have attracted lots
of experimental and theoretical interests \cite
{s1,s2,s3,s4,s5,s6,s7,s8,s9,s10,s11,s12,s13,s14,s15,s16,s17,s18}.
Despite the pairing mechanism being still controversial, a lot of
theoretical works indicate that the orbital degeneracy may play a
key role in these new family of high-Tc superconductors. Different
from the conventional cuperate superconductors where the undoped
compound is a Mott insulator, the pure LaOFeAs compound is a poor
metal. This implies that the orbital degeneracy may dramatically
affect the properties of the normal state. It is well known that
one of the most basic model to understand the strongly correlated
systems is the well-known single-band Hubbard model. Although
thousands of works have been focused on such a deceptively simple
model, the physical properties are not fully understood except the
one-dimensional (1D) case where the exact solutions are available
\cite{LiebW}. The exact results indicate that the Hubbard model
with a filling factor one is a Mott insulator at the zero
temperature. It is quite interesting to ask whether the 1D Hubbard
model with the same filling factor but with an additional orbital
degree of freedom is still a Mott insulator in the zero
temperature? Aiming to answer this problem, we study the
conductivity of an extended 1D Hubbard model with the orbital
degree of freedom in the scheme of the Bethe-ansatz solution.
Unlike the single-band Hubbard model where the conductivity is
found to be zero for any nonzero repulsive interactions, the
Hubbard model with the orbital degree is found to be a conductor
when the repulsive interaction is smaller than a critical value
and a phase transition from metal to an insulator occurs when the
on-site $U$ is larger than the critical value.

%\section{The model}

A 1D electronic system with the orbital degree of freedom can be modeled by
\begin{equation}
H=-t\sum_{i,a}\left( C_{i,a}^{+}C_{i+1,a}e^{i\phi /L}+h.c.\right)
+U\sum_{i,a<a^{\prime }}n_{i,a}n_{i,a^{\prime }},  \label{Hamiltonian}
\end{equation}
where $i=1,2,\dots, N$ identify the lattice site, $N$ is the total
particle number, $L$ is the system-size, and $a=1,2,...,4$ labels
the four states of single site, i.e. $\{1\uparrow, 1\downarrow,
2\uparrow, 2\downarrow\}$ with $1, 2$ being the indices of orbital
and $\uparrow, \downarrow$ being the indices of spin
\cite{LiMSZ98,LiGYE00}. The internal degree of freedom in the
Hamiltonian (\ref{Hamiltonian}) is specified to spin and orbital in
present model. The $C^+_{i a}$ creates an electron with spin-orbital
component $a$ on site $i$, and $n_{i a}:=C^{+}_{i a}C_{i a}$ is
the corresponding number operator at site $i$. The system (\ref{Hamiltonian}%
) is assumed with periodic boundary condition and $\phi$ is the
magnetic flux piercing the ring. The system (\ref{Hamiltonian}) is
the Hamiltonian for four-component systems, and there are various
discussions on
multi-component Hubbard model in one dimension \cite%
{LiGYE00,Choy,ChoyHaldane,Schlottmann,FrahmSS}.

The conductivity of a many-body system generally takes the form
\begin{equation}
\sigma (\omega )=2\pi D_{c}\delta (\omega )+\sigma _{r}(\omega ),
\end{equation}%
where $D_{c}$ is the charge stiffness and $\sigma _{r}(\omega )$\ is the
regular part of the conductivity. If $D_{c}$ is finite, the system is a
perfect conductor; and if $D_{c}$ is zero but $\sigma _{r}(\omega )$ is
finite, the system is a normal conductor; while if both of them are zero,
the system is an insulator. At zero temperature, the transport properties of
one-dimensional systems depend usually on the charge stiffness. Kohn showed
that the charge stiffness can be computed from the ground-state energy $%
E(\phi )$ as \cite{Kohn64}
\begin{equation}
D_{c}=\frac{N}{2}\frac{d^{2}E(\phi )}{d\phi ^{2}},
\end{equation}%
where $\phi $ is the external magnetic flux. It is well-known that
the magnetic flux $\phi $ piercing the system with the periodic
boundary condition can be gauged out by imposing the twisted
boundary conditions on the system \cite{su1,su2}. Therefore, solving
the Schrodinger equation in the presence of the magnetic flux with
periodic boundary condition is equivalent to that in the absence of
the magnetic flux but with a twisted boundary condition for the
wavefunctions
\begin{equation}
\psi (x_{1},\cdots ,x_{i}+L,\cdots )=e^{i\phi }\psi (x_{1},\cdots
,x_{i},\cdots ).
\end{equation}

We restrict our studies in the case of $L=N$. If $U=0$, the electrons do not
interact with each other and the Hamiltonian can be transformed as
\[
H=-2t\sum_{i,a}\cos (k+\phi /L)C_{k,a}^{+}C_{k,a}.
\]%
At the ground state, the electrons are arranged below the Fermi
surface according to the Pauli exclusive principle. Then the density
of electrons is
\[
\frac{N}{L}=\frac{1}{2\pi }\int_{-k_{F}}^{k_{F}}4dk,
\]%
where $k_{F}$ is the Fermi momentum and
\[
k_{F}=\frac{\pi N}{4L}.
\]%
For the case of $N=L$, the Fermi momentum is $k_F=\pi/4$. The
ground-state energy is
\begin{eqnarray}
\frac{E(\phi /L)}{L} &=&-\frac{4t}{\pi }\int_{-k_{F}}^{k_{F}}\cos (k+\phi
/L)dk, \\
E(\phi /L) &=&-\frac{4tL}{\pi }\left[ \sin (k_{F}+\phi /L)-\sin
(-k_{F}+\phi /L)\right].
\end{eqnarray}%
%where and in the following, the hoping $t$ is set to 1.
At the $U=0$ case, the ground-state charge stiffness can be obtained
analytically as
\[
D_{c}=\frac{2\sqrt{2}}{\pi }=0.9003163.
\]%
Therefore, the system is a perfect conductor. While if the
interactions between the electrons tends to infinity, $U\rightarrow
\infty $, each double occupation will cost an infinite energy thus
the each site favors the single occupation. The system is an
insulator. Therefore, a quantum phase transition from a conducting
phase to an insulating phase should occur between these two limiting
cases.

For finite $U$, the Hamiltonian is quasi-integrable if site occupations of
more than two electrons are excluded. Physically, this is reasonable for the
present studied case with filling factor one due to the state with more than
two electrons on a site is energy unfavorable. Following the standard
procedure \cite{Choy,ChoyHaldane,Schlottmann}, the energy of the system (\ref%
{Hamiltonian}) is
\begin{equation}
E(\phi )=-2t\sum_{j=1}^{N}\cos k_{j},
\end{equation}
where the quasi-momentum $k_j$ should satisfy following Bethe-ansatz
equations
\begin{eqnarray}
e^{ik_{j}L} &=&e^{i\phi }\prod_{b=1}^{M}\frac{\sin k_{j}-\lambda _{b}+i\eta
}{\sin k_{j}-\lambda _{b}-i\eta },  \nonumber \\
\prod_{l=1}^{N}\frac{\lambda _{a}-\sin k_{l}+i\eta }{\lambda _{a}-\sin
k_{l}-i\eta } &=&-\prod_{b=1}^{M}\frac{\lambda _{a}-\lambda _{b}+i2\eta }{%
\lambda _{a}-\lambda _{b}-i2\eta }\prod_{c=1}^{M^{\prime }}\frac{\mu
_{c}-\lambda _{a}+i\eta }{\mu _{c}-\lambda _{a}-i\eta },  \nonumber \\
\prod_{b=1}^{M}\frac{\mu _{a}-\lambda _{b}+i\eta }{\mu _{a}-\lambda
_{b}-i\eta } &=&-\prod_{c=1}^{M^{\prime }}\frac{\mu _{a}-\mu _{c}+i2\eta }{%
\mu _{a}-\mu _{c}-i2\eta }\prod_{d=1}^{M^{\prime \prime }}\frac{\nu _{d}-\mu
_{a}+i\eta }{\nu _{d}-\mu _{a}-i\eta },  \nonumber \\
\prod_{b=1}^{M^{\prime }}\frac{\nu _{a}-\mu _{b}+i\eta }{\nu _{a}-\mu
_{b}-i\eta } &=&-\prod_{c=1}^{M^{\prime \prime }}\frac{\nu _{a}-\mu
_{c}+i2\eta }{\nu _{a}-\mu _{c}-i2\eta },  \label{BAE}
\end{eqnarray}
where $\eta =U/4t$, $\lambda, \mu$ and $\nu$ are the rapidities.

\begin{figure}[h]
\begin{center}
\includegraphics[height=6cm,width=8cm]{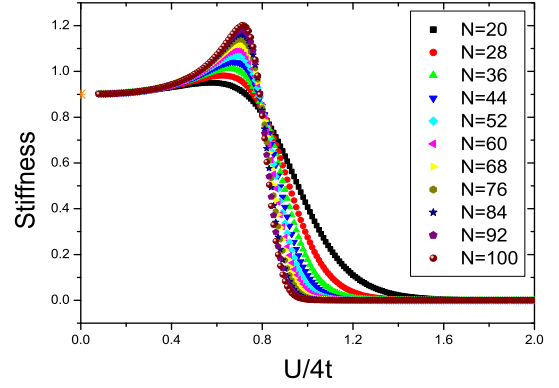}
\end{center}
\caption{(color online). The curve of charge stiffness versus the
on-site
couplings for $L=N$. The cross point on $y$-axis is $2\protect\sqrt{2}/%
\protect\pi $.} \label{fig1}
\end{figure}
For ground state (i.e., at zero temperature), the $k,\lambda ,\mu
,\nu $ are real roots of the Bethe ansatz equations (\ref{BAE}).
Taking the logarithm of the Bethe-ansatz equations, we get
\begin{eqnarray}
2\pi I_{j} &=&k_{j}L+\phi +2\sum_{a}\tan ^{-1}\left( \frac{\sin
k_{j}-\lambda _{a}}{\eta }\right) ,  \nonumber \\
2\pi J_{a} &=&2\sum_{l}\tan ^{-1}\left( \frac{\lambda _{a}-\sin k_{l}}{\eta }%
\right) -2\sum_{b}\tan ^{-1}\left( \frac{\lambda _{a}-\lambda _{b}}{2\eta }%
\right)  \nonumber \\
&&-2\sum_{c}\tan ^{-1}\left( \frac{\mu _{c}-\lambda _{a}}{\eta }\right) ,
\nonumber \\
2\pi K_{a} &=&2\sum_{b}\tan ^{-1}\left( \frac{\mu _{a}-\lambda _{b}}{\eta }%
\right) -2\sum_{c}\tan ^{-1}\left( \frac{\mu _{a}-\mu _{c}}{2\eta }\right)
\nonumber \\
&&-2\sum_{d}\tan ^{-1}\left( \frac{\nu _{d}-\mu _{a}}{\eta }\right) ,
\nonumber \\
2\pi Q_{a} &=&2\sum_{b}\tan ^{-1}\left( \frac{\nu _{a}-\mu _{b}}{\eta }%
\right) -2\sum_{c}\tan ^{-1}\left( \frac{\nu _{a}-\nu _{c}}{2\eta }\right),
\label{bae2}
\end{eqnarray}%
where $\{I_{j},J_{a},K_{a},Q_{a}\}$ are quantum numbers. $I_{j}$ takes
integer or half-odd integer depending on whether $M-1$ is odd or even. $%
J_{a},K_{a}$ and $Q_{a}$ take integer or half-odd integer depending on whether $%
N-M-M^{\prime }$, $M-M^{\prime }-M^{\prime \prime }$ and $M^{\prime
}-M^{\prime \prime }$ are integer or half-odd integers,
respectively. If $N=4n$ for $n$ being odd integer, the ground state
is non-degenerate, and quantum number are centerred symmetrily
around the zero point.

%\section{Conductivity}
\begin{figure}[h]
\begin{center}
\includegraphics[height=6cm,width=8cm]{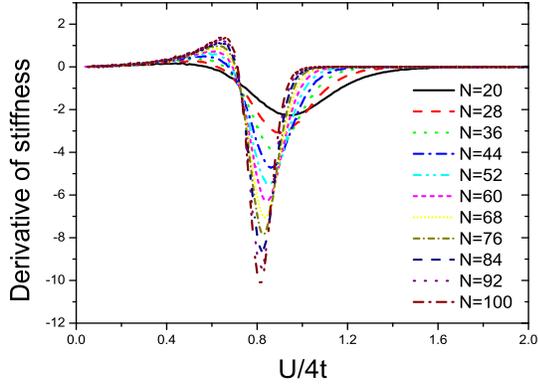}
\end{center}
\caption{(color online). The derivative of the charge stiffness. At
some certain coupling $U_{m}$, the derivative has a minimum.}
\label{fig2}
\end{figure}
\begin{figure}[h]
\begin{center}
\includegraphics[height=6cm,width=8cm]{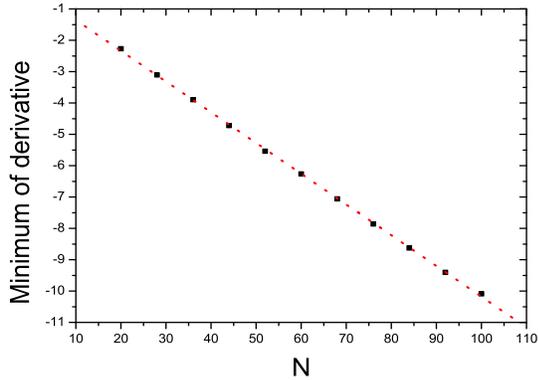}
\end{center}
\caption{(color online). The curve of the minimum of derivative of the stiffness $%
D^{\prime}_m$ versus the system size $N$. The data can be fitted as $%
D^{\prime}_m = -0.374 -0.098 N$. One see that the minimum is
divergence if the system size tends to infinity.} \label{fig3}
\end{figure}
\begin{figure}[h]
\begin{center}
\includegraphics[height=6cm,width=8cm]{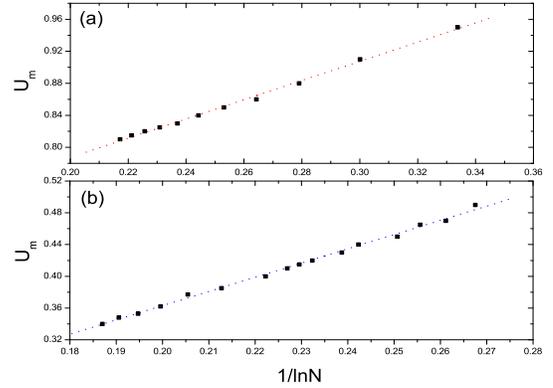}
\end{center}
\caption{(color online). (a) The scaling behavior of the $U_m$. The
data of the two-band model can be fitted as $U_m/4t=0.5475+
1.19993/\ln N$. When the lattice
number $N$ tends to infinity, $U_m$ becomes $U_c$, the critical value of $%
U_c/4t$ reads $0.548\pm 0.005$. (b) The same scaling analysis is
performed for the single band Hubbard model. The data of the single
band model can be fitted as $U_m/4t=0.00483 + 1.79084/\ln N$. The
critical value of $U_c/4t$ reads $0.005\pm 0.006$, which covers the
exact critical point $U_c=0$ of the single band Hubbard model.}
\label{fig4}
\end{figure}
\begin{figure}[h]
\begin{center}
\includegraphics[height=6cm,width=8cm]{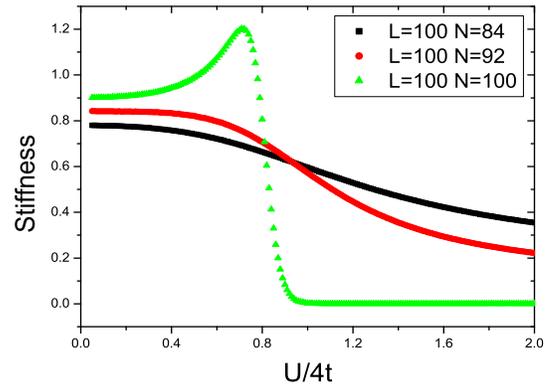}
\end{center}
\caption{(color online). Charge stiffness with holes. Here the
system size $L=100$ and the particle numbers $N=84, 92, 100$. The
charge stiffness has a sharp peak only at the case of $L=N$. }
\label{fig5}
\end{figure}
We numerically solve the Bethe ansatz equations (\ref{bae2}) with
the finite system-size $L$. The charge stiffness versus the
interaction $U$ is shown in Fig. \ref{fig1}. We see that the charge
stiffness shows a sharp peak with the increasing system size. In the
thermodynamic limit, the charge stiffness is expected as a step
function, which takes a non-zero value at one side and zero at
another side. The sudden jump point defines the critical $U_{c}$ of
the phase transition. For the present model, if $U<U_{c}$, the
system is a metal while if $U>U_{c}$, the system is an insulator.
While for the single band Hubbard model, Lieb and Wu show that the
Mott-insulator transition only happens at the $U_{c}=0$ case.

The critical $U_{c}$ can be determined by the derivative of the
charge stiffness. The derivative of the charge stiffness versus the
coupling $U$ is shown in Fig. \ref{fig2}. We see that the derivative
has a minimum at a certain $U_{m}$. The minimum is decreasing with
the increasing system-size. When the system-size tends to infinity,
the minimum is divergent, which can be seen clearly in Fig.
\ref{fig3}. From Fig. \ref{fig3}, the value of the minimum of
derivative of the charge stiffness versus the system-size $N$ can be
fitted into a straight line. Thus the charge stiffness becomes
steeper and steeper as the system size increases.

The critical coupling $U_{m}$ with finite system-size versus the
system-size is shown in Fig. \ref{fig4}. The data of $U_{m}$ and
system-size $N(=L)$ can be fitted as $U_{m}/4t=0.5475+1.19993/\ln
N$. When the system-size tends to infinity, $U_{m}$ becomes $U_{c}$.
The critical $U_{c}/4t$ reads $0.548\pm 0.005$. We also perform the
same scaling analysis for the single-band Hubbard model [Fig.
\ref{fig4}(b)] and find that $U_{c}/4t=0.005\pm 0.006$. The
difference between the two models is clear. For the case of $N=L$,
the multiband might enhance the critical value $U_{c}/4t$.

Now we consider the system with some holes. For this case the Bethe
ansatz solutions are solved by choosing suitable quantum numbers
$\{I_{j}, J_{a}, K_{a}, Q_{a}\}$. The charge stiffness versus the
coupling $U$ is shown in Fig. \ref{fig5}, where the system-size is
set to $L=100$ and particle numbers are $N=84, 92, 100$. We see that
only at the case of $N=100$, the curve has a sharp transition; while
for both cases $N=84$ and $N=92$, the charge stiffness always takes
a nonzero value. This observation is consistent with the fact that
the system would be a metal if we add some holes.

%\section{Conclusion}

In conclusion, starting from the Bethe ansatz solutions of 1D
two-band Hubbard model, we study the multiband effects on the
conductivity. We find that multiband would enhance the critical
value $U_{c}$ of on-site interactions of the transition from a metal
to an insulator, while the critical $U_{c}$ for the single band
Hubbard model is zero. The finite system-size would have a $1/\ln L$
correction to the actual value. The orbital degree of freedom might
play an essential role in the properties of the electronic systems.
These results are helpful to understand why undoped High-$T_{c}$
superconductors are usually insulators, while recently discovered
iron-based superconductors are metal even without doping.

%\section*{Acknowledgments}

This work was supported by the Earmarked Grant for Research from the
Research Grants Council of HKSAR, China (Projects No.HKU\_3/05C),
the national natural science foundation of China, and the national
program for basic research of MOST. S. J. Gu is grateful for the
hospitality of Institute of Physics at Chinese Academy Sciences.


\begin{thebibliography}{99}
\bibitem{s1} Y. Kamihara, T. Watanabe, M. Hirano and H. Hosono, J. Am. Chem.
Soc. \textbf{130}, 3296 (2008).

\bibitem{s2} D. J. Singh and M. H. Du, cond-mat/0803.0429.

\bibitem{s3} K. Haule, J. H. Shim and G. Kotliar, cond-mat/0803.1279.

\bibitem{s4} G. Xu, W. Ming, Y. Yao, X. Dai, S. Zhang and Z. Fang,
con-mat/0803.1282.

\bibitem{s5} C. Cao, P. J. Hirschfeld and H. P. Cheng, cond-mat/0803.3236.

\bibitem{s6} X. H. Chen, T. Wu, G. Wu, R. H. Liu, H. Chen and D. F. Fang,
cond-mat/0803.3603.

\bibitem{s7} G. F. Chen, Z. Li, D. Wu, G. Li, W. Z. Hu, J. Dong, P. Zheng,
J. L. Luo, N. L. Wang, cond-mat/0803.3790.

\bibitem{s9}Z. A. Ren, et. al.,  Chin. Phys. Lett. 25, 2215 (2008)
; Z. A. Ren, et. al., Europhys. Lett., 82 (2008) 57002.

\bibitem{s16} H. H. Wen, G. Mu, L. Fang, H. Yang and X. Zhu, Europhys. Lett.
\textbf{82}, 17009 (2008).

\bibitem{s12} Z. A. Ren, G. C. Che, X. L. Dong, J. Yang, W. Lu, W. Yi, X. L.
Shen, Z. C. Li, L. L. Sun, F. Zhou and Z. X. Zhao,
cond-mat/0804.2582.

\bibitem{s8} X. Dai, Z. Fang, Y. Zhou and F. C. Zhang, cond-mat/0803.3982.

\bibitem{s10} P. A. Lee and X. G. Wen, cond-mat/0804.1739.

\bibitem{s11} Q. M. Si and E. Abrahams, cond-mat/0804.2480.

\bibitem{s13} S. Ishibashi, K. Terakura and H. Hosono, cond-mat/0804.2963.

\bibitem{s14} F. Ma, Z. Y. Lu and T. Xiang, cond-mat/0804.3370.

\bibitem{s15} Z. J. Yao, J. X. Li and Z. D. Wang, cond-mat/0804.4166.


\bibitem{s17} X. L. Qi, S. Raghu, C. X. Liu, D. J. Scalapino and S. C.
Zhang, cond-mat/0804.4332.

\bibitem{s18} J. Li and Y. Wang, Chin. Phys. Lett. \textbf{25}, 2232 (2008).

\bibitem{LiebW} E. H. Lieb and F.Y. Wu, Phys. Rev. Lett. \textbf{25}, 1445
(1968).

\bibitem{LiMSZ98} Y. Q. Li, M. Ma, D. N. Shi, and F. C. Zhang, Phys. Rev.
Lett. \textbf{81}, 3527 (1998).

\bibitem{LiGYE00} Y. Q. Li, S. J. Gu, Z. J. Ying, and U. Eckern, Phys. Rev.
\textbf{B62}, 4866 (2000).

\bibitem{Choy} T. C. Choy, Phys. Lett. \textbf{80 A}, 49 (1980).

\bibitem{ChoyHaldane} T. C. Choy and F. D. M. Haldane, Phys. Lett. A \textbf{%
90}, 83 (1982).

\bibitem{Schlottmann} P. Schlotmann, Phys. Rev. \textbf{B43}, 3101 (1991).

\bibitem{FrahmSS} H. Frahm and A. Schadschneider, \textit{The Hubbard model:
Its Physics and Mathematical Physics} Eds. D. Baeriswyl et al., D. K.
Campbell, J. M. P. Carmelo, F. Guinea, and E. Louis (Plenum Press, New York
1995) pp. 21; P. Schlottmann, Int. J. Mod. Phys. \textbf{B} 11, 355 (1997).


\bibitem{Kohn64} W. Kohn, Phys. Rev. \textbf{133}, A171 (1964).

\bibitem{su1} N. Byers and C.N. Yang, Phys. Rev. Lett. \textbf{7}, 46 (1986).

\bibitem{su2} B. S. Shastry and B. Sutherland, Phys. Rev. Lett. \textbf{65},
243 (1990).

\end{thebibliography}
\end{document}